\documentclass[nofootinbib,preprintnumbers,floatfix]{revtex4}

\usepackage{color}
\usepackage{graphicx}
\usepackage{epstopdf}
\usepackage{amsmath}
\usepackage{amssymb}
\usepackage{xspace}
\usepackage{epsfig}
\usepackage[small]{subfigure}
\usepackage[hyperfootnotes=false]{hyperref}
\usepackage{pstricks}

\newcommand{\eq}[1]{Eq.~\eqref{eq:#1}}
\newcommand{\eqs}[2]{Eqs.~\eqref{eq:#1} and \eqref{eq:#2}}
\renewcommand{\sec}[1]{Sec.~\ref{sec:#1}}

\newcommand{\fig}[1]{Fig.~\ref{fig:#1}}

\newcommand{\df}{\mathrm{d}}

\newcommand{\tr}{\textrm{tr}}

\newcommand{\cP}{{\mathcal P}}

\newcommand{\bn}{{\bar{n}}}

\newcommand{\nn}{\nonumber}

\newcommand{\mcdot}{\!\cdot\!}

\newcommand{\s}{{s}}

\newcommand{\bCH}[2]{\overline\chi_{#1,#2}}

\allowdisplaybreaks[2]

\color{black}

\begin{document}


\preprint{\vbox{\hbox{MIT--CTP 4102}\hbox{arXiv:0912.5509}}}

\title{Jet Function with a Jet Algorithm in SCET}

\author{Teppo T.~Jouttenus}
\affiliation{Center for Theoretical Physics, Massachusetts Institute of
  Technology, Cambridge, MA 02139\vspace{2ex}}

\begin{abstract}
  The jet function for the factorized cross section $e^+e^-$ into dijets is
  given as a function of the jet invariant mass $\s$ and with a generic jet
  algorithm at $\mathcal{O}(\alpha_s)$. We demonstrate the results using the
  Sterman-Weinberg algorithm and show that the jet function is independent of
  the energy fraction $\beta$ of the soft radiation. The anomalous
  dimension has the same form with and without the cone half-angle $\delta$. The
  dependence of the finite part of the jet function on the cone angle is given.

\end{abstract}

\maketitle

\section{Introduction}
\label{sec:intro}

Hadronic jets feature in many final states of interest in modern collider experiments. They form a significant Standard Model background for many new physics processes and also provide probes for QCD interactions at several different scales. In order to disentangle the effects of these different momentum scales, factorization is needed. Factorization theorems make it possible to separate process-dependent perturbative physics from universal nonperturbative effects. Soft-collinear effective theory (SCET) \cite{Bauer:2000ew, Bauer:2000yr, Bauer:2001ct, Bauer:2001yt, Bauer:2002nz} provides a framework for deriving factorization theorems while systematically resumming large logarithms to all orders in perturbation theory and including power corrections to any desired accuracy. The power expansion is performed in terms of a parameter \(\lambda\), which characterizes the ratio of the transverse and the collinear momentum in a jet. 

The cross section for $ e^+e^- $ into dijets  can be factorized schematically as follows
\begin{equation}
\label  {eq:factorize}
        \frac{1}{\sigma_0}\frac{\df \sigma}{\df s \, \df \bar{s}} =  H ( J_  {n} \otimes J_  { \bn} \otimes S),
\end{equation}
where  $ \s$ and $\bar{\s} $ are the invariant masses of the jets, $ \sigma_0 $ is the Born cross section,  $ H $ is the hard coefficient obtained by matching SCET to QCD,  $  J_{n/ \bn}  $ are the jet functions in $ n$ and $\bn $ directions,   $ S $ is the soft function, and the convolution is in the small light-cone momentum component for each jet. We study a cross section differential in $\s$ and $\bar{\s}$ because after specifying the total energy and the direction of a jet, the invariant mass is the next natural observable to consider in order to find out more about the structure of the jet. 
Factorization theorems for $ e^+e^- $ colliders have been derived in SCET for event shape variables using a hemisphere jet algorithm \cite{Fleming:2007xt, Hornig:2009vb}. For hadron colliders, factorization theorems have been studied at a general level \cite{Bauer:2008jx} as well as for threshold resummation \cite{
Sterman:1987,
Catani:1989,
Kidonakis:1998bk,
Bonciani:1998vc,
Laenen:1998qw,
Catani:2003zt,
Idilbi:2006dg,
Becher:2007ty,
Chiu:2008vv} and for isolated Drell-Yan \cite{Stewart:2009yx} defined as having no central jets.

To match experimental results more precisely, a theory calculation should define
jets using the same jet algorithm that is used in a given experimental analysis.
The present work is a first step towards the final goal of proving a
factorization theorem for the dijet cross section given above with a realistic
jet algorithm. Instead of attacking the problem all at once, we derive a
formalism to calculate the complete jet function to \( \mathcal{O}(\alpha_s) \)
and to leading order in power counting. Our results apply for any jet algorithm
that can be formulated in terms of theta functions depending on the momenta of
the final state particles and that can be regularized using dimensional
regularization. As an example, we look at the Sterman-Weinberg algorithm (SW)
\cite{Sterman:1977wj}, which has been considered using SCET in
Ref.~\cite{Bauer:2002ie, Bauer:2003di, Trott:2006bk, Cheung:2009sg}. We
calculate the full jet function \( J_{n} \), including the finite part, as a
function of the invariant mass of the jet. After an expression for the soft
function has been calculated, our result can be used to derive a factorized
cross section $e^+e^-$ into dijets with the jets defined using the SW algorithm.

The SW algorithm defines a dijet event as one where all but a fraction \(\beta  \) of the total energy is contained within a pair of oppositely directed cones of half-angle \(\delta \) \cite{Sterman:1977wj}. The definition can also be extended for more jets if one specifies a way to determine the direction of the different jet axes. Very recent work by
Ellis et al.~\cite{Ellis:2009wj} shows that in the case of more than two jets, consistent factorization requires the jets to be well separated and the radiation outside the jets to be soft. Hence, we expect a factorization proof for  $N$ jets defined with SW algorithm to take the schematic form
\begin{equation}
\label  {eq:factorizeSW}
        \frac{1}{\sigma_0}\frac{\df \sigma^{ \textrm{SW}}(\delta,\beta)}{\df s_1 \ldots \df s_N} =  H \left(  \prod^N_{i=1} J^  {\rm SW}_  {n_i}  \otimes S^  {\rm SW}\right),
\end{equation}
with $\mathcal{O}(\delta, \beta)$ power corrections. For the expected power corrections to be subleading in power counting, it is natural to take $\delta, \beta \sim \lambda$. As will be seen later, to leading order in $\beta$ the final result for the jet function is independent of $\beta$, even if we pick $\delta \sim \lambda, \ \beta \sim \lambda^2$. It would seem logical to also consider $\delta \sim \lambda^2, \ \beta \sim \lambda$ but as has been pointed out in Ref.~\cite{Stevenson:1978td}, one should require  $ \sin (\delta) > \beta / (1-\beta) $ in order to preserve the back-to-back orientation of the two jets. To lowest order in the parameters, this corresponds to $\delta > \beta$. Finally, one could  consider $\delta, \beta \sim \lambda^2$ but then it would be more natural to define a new expansion parameter $\lambda' \equiv \lambda^2$ and set $\delta, \beta \sim \lambda'$. Thus, the natural possibilities to consider are $\delta \sim \beta \sim \lambda$ or $\delta \sim \lambda$, $\beta \sim \lambda^2$ and our results for the jet function are valid for both cases. For the sake of definiteness, we will take  $ \delta \sim \beta \sim \lambda $ with the understanding that $ \sin (\delta) > \beta / (1-\beta) $. 

Ref.~\cite{Ellis:2009wj} also provides a useful cross check for our results. They have demonstrated the consistency of
factorization of jet observables in exclusive multijet cross sections for both cone and cluster type algorithms and given
expressions for the anomalous dimensions of the hard, jet, and soft functions. The present work agrees with their
result for the anomalous dimension of the jet function. We also derive the
finite contribution, which has not been given in the literature
before.
\footnote{Right after this paper appeared, a second paper by Ellis et
  al.~\cite{Ellis:2010rw} came out. It is discussed in a note added and an
  Appendix at the end.  }

In \sec{genericalg} we review the necessary elements of SCET and present our
results for a generic jet algorithm. In \sec{SWalg} we express the SW jet
definition in terms of the components of the momenta of the final state
particles. In \sec{calculations} we present the main results of our calculation.
We plot the renormalized jet function both as a function of $\s$ and integrated
over $\s$. For comparison, the jet function is also plotted without a jet
algorithm. We conclude in \sec{conclusions}.

\section{Jet Function with Generic Algorithm}
\label{sec:genericalg}

Our goal is to derive a formalism to accommodate any jet algorithm that can be expressed in terms of phase space cuts. In this section we outline the derivation for an expression for the jet function \( J_{n}(\s) \). We start by reviewing some conventions for notation. The direction of the jet axis is denoted  by a unit vector  $ \vec  { n} $. We choose a coordinate system such that  $ \vec  { n}=(0,0,1) $. It is convenient to work in the light cone coordinates with basis vectors  $ n^\mu = (1,\vec  { n} ) $ and $ \bn^\mu = (1,-\vec  { n} ) $ satisfying  $ n^2 = \bn^2 = 0, \  n \mcdot \bn = 2$. Then any momentum can be decomposed as  
\begin{eqnarray}
p^\mu = n \mcdot p \>\frac{\bn^\mu}{2} + \bn \mcdot p\> \frac{n^\mu}{2} +
p_\perp^\mu \equiv \left( n \mcdot p, \, \bn \mcdot p,\, p_\perp \right) \equiv \left( p^+, \,  p^-,\, p_\perp \right).
\end{eqnarray}
The momentum of a collinear parton $i$ in  $ n $-direction scales as $p_i \sim Q ( \lambda^2, 1, \lambda)$ so we can write  $ p^0_i$, $p^3_i = \frac{1}{2} p^-_i + \mathcal{O}( \lambda) $. The center-of-mass energy is denoted by $Q$ and the total jet momentum is $ p_n+r_n = (Q,r^+,0) $, where  $ p_n $ is the large label momentum and  $ r_n $ is the smaller residual momentum. Thus, the invariant mass of the jet is given by $ \s \equiv (p_n+r_n)^2 = Q r^+_n \sim Q^2 \lambda^2$. Because of the specific observable we consider, namely the jet invariant mass, we are free to use the coordinate system specified above, which removes all dependence on the total transverse momentum $p^\perp_n$. For a general observable, we would also need to include $p^\perp_n$. To define a jet algorithm, we also need to consider the momenta of the final state partons. At \( \mathcal{O}(\alpha_s) \), momentum conservation and the on-shell condition for the final state partons make it possible to write all the momenta in the problem in terms of the jet momentum components $p_n^- = Q$ and $r_n^+$ together with the gluon four-momentum $p_g$. 

The jet function can be written in terms of the gauge invariant quark jet field $\chi_{n }= W^\dag_n \, \xi_n$, and it is also convenient to use $ \chi_{n, \omega }= \delta( \omega- \bn \cdot \cP) (W^\dag_n \xi_n)$, where  $ \bn \cdot \cP $ gives the large label momentum of the combination  $  W^\dag_n \xi_n $. The collinear Wilson line is defined as
\begin{eqnarray}
 W_n (x) = \sum _{\text{perms}} \text{exp} \bigg[- \frac{g}{ \bn \mcdot \cP }
   \bn \mcdot A_{n}(x) \> \bigg],
\end{eqnarray}
where  $ A_n $ is the collinear gluon field. 

We will use the term inclusive jet function to refer to the case where no jet algorithm is applied and denote it by $J^{\rm (inc)}_{n} $. It is discussed in Ref.~\cite{Bauer:2003pi, Becher:2006qw} and we define it as
\begin{align}
\label{eq:jetfunc}
J^{\rm (inc)}_{n}(\s,\mu) 
        &= 
        \frac{-1}{8\pi N_c Q } \, \textrm{Disc} \! \int \!\! \df^d x \, 
        e^{i r_n\cdot x} \,
        \tr \langle 0|\text{T}\{ \bCH n Q (0)\slash\!\!\!\bn  \chi_n(x)\}|0 \rangle
        \notag \\
        &= 
        \frac{1}{8\pi N_c Q }  \sum_{X_{n}} \int \!\! \df^d x \, 
        e^{i r_n\cdot x} \,
        \tr \langle 0| \slash\!\!\!\bn  \chi_n(x) |X_{n} \rangle \langle X_{n} |  \bCH n Q (0)|0 \rangle ,
\end{align}
where $ N_c $ is the number of colors, $ d= 4 - 2 \epsilon $,  the trace is over color and spin, and  $ \text  {T} $ stands for time ordering. Because of charge conjugation symmetry, the antiquark jet function \( J_{\bn} \) does not have to be considered separately. We write \( J_{n} \) as a sum over final states as in Ref.~\cite{Fleming:2007xt} in order to implement a jet algorithm. The final states are restricted according to a constraint function $F(a_i)$, which defines an algorithm in terms of parameters  $ a_i $ and depends on the momenta of the particles in $|X_{n} \rangle $. Thus, $F(a_i)$ is also a function of the operators $ \hat{p}_j $ which have the final state momenta  $ p_j $ as eigenvalues. In this section, we will work with a generic  $ F(a_i, \hat{p}_j)$ and in \sec  {SWalg} we will specialize to the SW algorithm. Inserting the constraint function gives the algorithm-dependent jet function
\begin{figure}
       \begin{center}
       \includegraphics[width=6.0in]{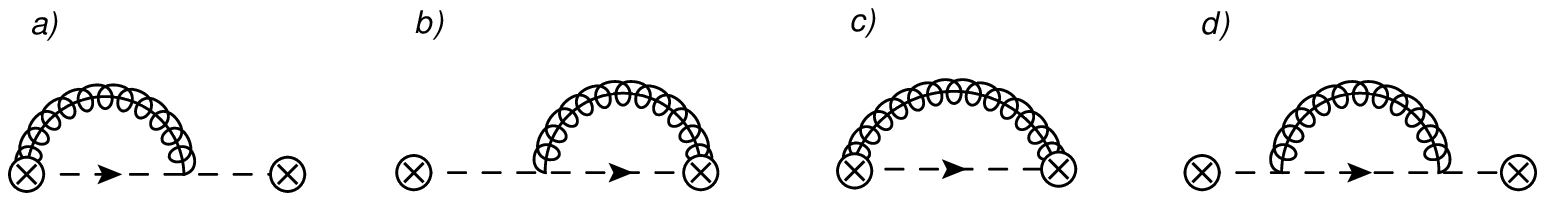} 
       \end{center}
       \begin{center}  
       \includegraphics[width=6.0in]{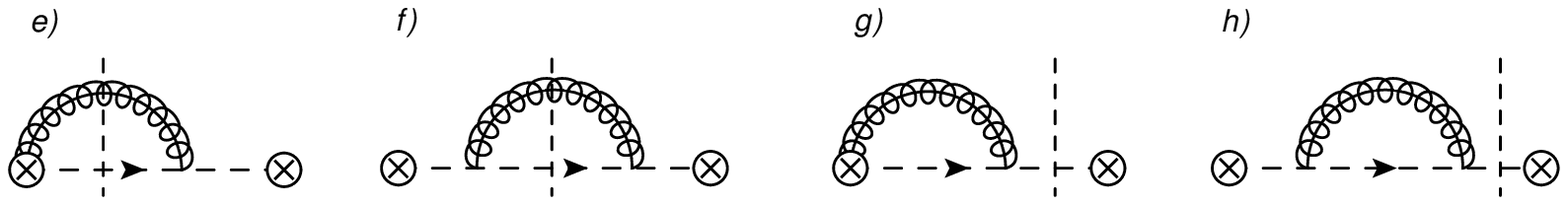}
       \caption{The Wilson line diagrams (a) and (b) give identical contributions. Diagram (c) vanishes in Feynman gauge. Diagram (d) gives the wavefunction renormalization contribution. Diagrams (e) and (f) show the real cuts and diagrams (g) and (h) show the virtual cuts. The mirror image of (h) gives the other virtual cut of (d).}
       \label{fig:feyndiag}
       \end{center}
\end{figure} 
\begin{align}
\label{eq:genalgjetfunc}
J^{F}_{n}(\s,\mu ) 
        &= 
        \frac{1}{8\pi N_c Q }  \sum_{X_{n}} \int \!\! \df^d x \, 
        e^{i r_n\cdot x} \,
        \tr \langle 0| \slash\!\!\!\bn  \chi_n(x)F(a_i, \hat{p}_j) |X_{n} \rangle \langle X_{n} |  \bCH n Q (0)|0 \rangle.
\end{align}
\eq{genalgjetfunc} is valid at any order in \(\alpha_s\) but the functional form of $F(a_i,\hat{p}_j )$ changes from order to order.
Following the discussion of Hornig et al.~on angularity jet functions in Ref.~\cite{Hornig:2009vb}, we implement the needed phase space restrictions on the final states by introducing an $F$-discontinuity, where the standard Cutkosky cutting rules are modified by inserting a factor of $ F(a_i, \hat{p}_j) $ into the cut propagators. The diagrams that contribute to \( J^F_{n} \) at one loop are shown in \fig  {feyndiag} (a) - (d). We note that diagram (c) vanishes in Feynman gauge. At this order, we can cut through the loop or a single quark propagator, which we call ``real" and ``virtual" cuts, respectively. Diagrams (e) and (f) give the real cuts and (g) and (h) show the virtual cuts. The virtual cuts are independent of the algorithm and are contained in a proper interpretation of the inclusive jet function. Hence, the algorithm-dependent contribution is given by taking the real cuts and inserting $ F(a_i, \hat{p}_j) $ to the cut propagators. In order to use the known results from the literature more conveniently, we add and subtract the inclusive jet function on the second line.
\begin{align}
\label{eq:algjetfunc}
J^F_{n}(\s) 
     &=  \frac{-1}{8\pi N_c Q }  \textrm{Disc}_F\! \int \!\! \df^d x \, 
        e^{i r_n\cdot x} \,
        \tr \langle 0|\text{T}\{ \bCH n Q (0)   \slash\!\!\!\bn  \chi_n(x)\}|0 \rangle   \nn \\
        &= J^  {\rm (inc)}_{n}(\s)     +
        \frac{-1}{8\pi N_c Q }  \textrm{Disc} \! \int \!\! \df^d x \, 
        e^{i r_n\cdot x} \,
        \tr \langle 0|\text{T}\{ \bCH n Q (0)\left( F(a_i, \hat{p}_j) - 1\right)  \slash\!\!\!\bn  \chi_n(x)\}|0 \rangle 
        \equiv J^  {\rm (inc)}_{n}(\s) +\Delta J^F_{n},
\end{align} 
where we have introduced notation $ \Delta J^F_{n}$ for the algorithm dependent contribution. Note that setting $F=1$ in \eq{algjetfunc} leads to $\Delta J^F_{n}=0$ leaving only the inclusive contribution, as expected. To find an expression for  \( \Delta J^F_{n} \), we use the modified cutting rules to add a factor of $F(a_i, \hat{p}_j)$ and to replace the gluon and quark propagators in the loop by delta functions $\delta \left(p^-_g p^+_g + \left( p^\perp_g  \right)^2 \right)$ and $\delta \left(p^-_q p^+_q + \left( p^\perp_q  \right)^2 \right)$. Using the momentum conservation relation $p_q + p_g = (Q,r^+,0)$ and the above
delta functions, the phase space integrals over the gluon momentum components
$p_g^+$ and $p_g^\perp$ can be performed to give the relations
\begin{align} \label  {eq:deltaconditions}
 p^+_g   = -\frac{ \left( p^\perp_g  \right)^2}{ p^-_g}, \qquad \left( p^\perp_g  \right)^2 = -\frac{p^-_g ( Q - p^-_g) \s}{ Q^2}.
\end{align}
These relations must be used when writing the constraint function. Finally, the calculations can be simplified by a change of variables $y \equiv p_g^- / Q$. 

In a factorization theorem the contribution of soft quarks and gluons is encoded in the soft function. In order to avoid double-counting when the loop momentum of a collinear field in the jet function becomes soft, a zero-bin subtraction must be performed \cite{Manohar:2006nz}. The naive collinear result without the subtraction is obtained by summing over the contributions from the real cut Feynman diagrams in \fig{feyndiag}, diagrams (e) and (f), (and counting diagram (e) twice to account for its mirror image). This gives
\begin{align}
\label{eq:jetfunccoll}
        \Delta & \tilde{J}^F_{n} (\s) 
         =\frac{  \alpha_s C_F}{ 4 \pi }A( \epsilon) 
          \frac{1}{\mu^ 2} \left( \frac{\mu ^2}{s} \right)^{1+ \epsilon} \!
         \int^1_0 \! \df y \ \frac{1}{y^ \epsilon}(1-y)^{-\epsilon}  \left( \frac{4 (1-y)}{ y} +  y (d-2)  \right) (F(a_i, y)-1),
\end{align}
where \eq{deltaconditions} has been used to write $F$ as a function of $ y $ and
the algorithm parameters $ a_i $, the tilde denotes that the zero-bin
subtraction has not been performed, and $A( \epsilon) = 1 - \frac{ \pi^2
  \epsilon^2}{12} + \mathcal{O}( \epsilon^3)$. For the SW algorithm, we will see
that after the zero-bin has been subtracted, \( \Delta J_{n} \) is finite as $
\epsilon \rightarrow 0$ so we will eventually take $A( \epsilon) = 1$. There is
a zero-bin contribution both for the gluon and the quark becoming soft but only
the former contributes at leading order in power counting. Furthermore, as the
soft gluon wavefunction renormalization vanishes in Feynman gauge, we only need
to consider a zero-bin for the gluon in \fig{feyndiag} (e). To obtain the
zero-bin result, we assign scaling $p_g \sim Q\lambda^2$ \cite{Manohar:2006nz}
to all components of the gluon momentum. The effect of the zero-bin scaling to
algorithm constraints is discussed in detail in Appendix \ref{sec:zbforalg}. The
end result is that instead of \eq{deltaconditions}, the replacement rules for
the zero-bin piece are 
\begin{align} \label  {eq:ZBdeltaconditions}
 p^+_g   = \frac{\s}{ Q}, \qquad \left( p^\perp_g  \right)^2 = -p^-_g p^+_g = -\frac{p^-_g  \s}{ Q}, \qquad p^-_q =Q, \qquad p^+_q = 0, \qquad  p^\perp_q  = 0.
\end{align}
Using these relations, the zero-bin contribution to \( \Delta J^F_{n} \) is
\begin{align}
\label{eq:jetfuncZB}
        \Delta & J^F_{n0} (\s) 
         =\frac{  \alpha_s C_F}{ 4 \pi } A( \epsilon)
          \frac{1}{\mu^ 2} \left( \frac{\mu ^2}{s} \right)^{1+ \epsilon} \!
         \int^\infty_0 \! \df y \ \frac{4}{y^{1+ \epsilon}}  (F_0(a_i, y)-1),
\end{align}
where $F_0(a_i, y) $ denotes that the constraints are written using
\eq{ZBdeltaconditions}. For a very inclusive jet algorithm such as the
hemisphere algorithm, the zero-bin contribution remains a scaleless integral
that vanishes in dimensional regularization. However, a more restrictive
algorithm can introduce a scale that gives a nontrivial zero-bin subtraction and
we will see that this is what happens for the SW case.

\section{Sterman-Weinberg algorithm}
\label{sec:SWalg}

As explained in the introduction, the SW algorithm defines a dijet event in terms of the cone half-angle \(\delta \) and energy fraction of the soft radiation  \(\beta  \).  We will keep only the leading order in  \(  \delta \sim \beta \sim \lambda\), including both power and logarithmic dependence. When we apply the SW algorithm to the jet function, the question at one loop is whether the  $ n $-collinear quark and gluon create one or two jets. If both partons lie within \(\delta\) from the jet axis or if one of the partons has energy  $ E_i  < \beta \, Q$, only a single jet is produced. As stated before, a soft quark would contribute only at higher order in the power counting so we have two regions of phase space to consider: a ``cone'' region where both particles lie inside the cone and an ``outside-cone'' region where the gluon is not inside the cone and has energy $ E_g  < \beta \,Q$. We take the jet axis to lie along the  $ \vec  { n} $-direction and then the cone region corresponds to the momenta of both partons lying within \(\delta\) of $ \vec  { n} $.

The invariant mass $ \s $ of the jet is related to how far apart the final state
particles are spread. At one loop and using the on-shell condition for massless
partons, we find that $\s = p^-_g p^-_q \sin^2 (\phi/2) \approx p^-_g p^-_q
(\phi/2)^2$ where $ \phi $ is the angle between the quark and the gluon. If no
jet algorithm is applied, in the center-of-mass frame momentum conservation
implies $\s \le \left(Q /2 \right)^2$. In the SW algorithm with the gluon inside
the cone, the maximum value is reached when $ p^-_g = p^-_q = Q/2 $ and $ \phi =
2 \delta $, which gives $\s \le \left(Q \delta/2 \right)^2$. This constraint is
satisfied by the naive contribution to the jet function which only has support
for $\s \le \left(Q \delta/2 \right)^2$. However, the standard approach to the
gluon zero-bin subtraction is to utilize the result obtained in the zero-bin
limit everywhere, and hence there is no upper limit on $ p^-_g$ or
on $\s$. It may be possible to use the freedom in defining a zero-bin
subtraction scheme to define a non-minimal subtraction where one would
obtain a jet function which is nonzero only for $\s \le \left(Q \delta/2
\right)^2$. We leave this question to future work.  When the gluon is outside
the cone, the maximum is reached when $ y=2 \beta $ and $ \phi = \pi $, which
gives $ \s \le 2 \beta \, Q^2 $.  However, because of the power counting $\s
\sim Q \lambda^2$, this condition is always satisfied to leading order.

\begin{figure} 
\begin{center}
\includegraphics[width=3in]{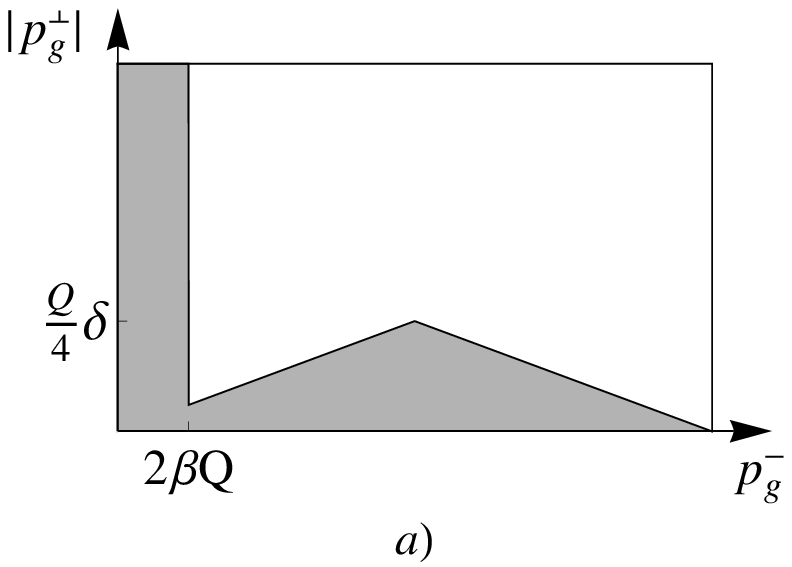}
\includegraphics[width=3in]{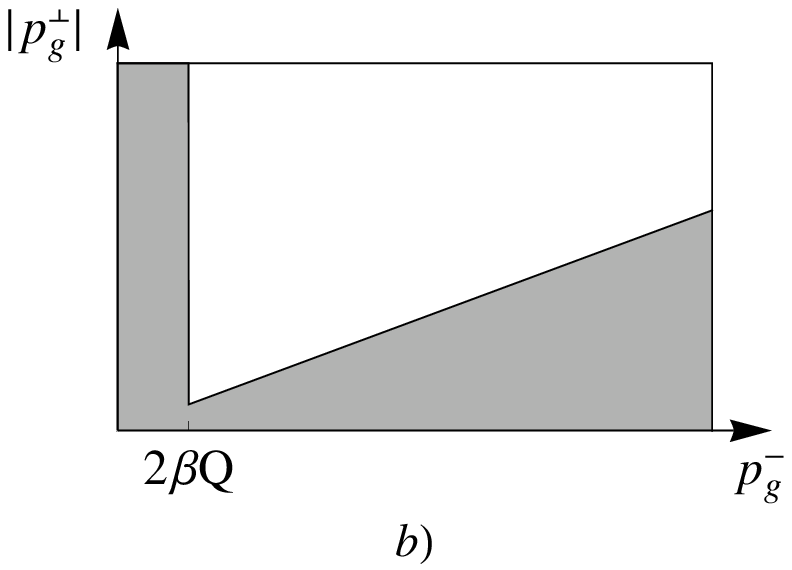}
\caption{Phase space regions for the SW algorithm. Naive collinear (a) and zero-bin (b). The region at the left-hand side of the figures corresponds to a gluon outside the cone with energy $E_g \le \beta Q$. The region at the bottom of the figures describes the region where both partons are inside the cone. For the zero-bin contribution, $p^-_g$ is integrated all the way to infinity. }
\label{fig:regions}
\end{center}
\end{figure}

The constraints for the SW algorithm are illustrated in \fig{regions} in terms of the variables $p^-_g$ and $p^\perp_g$.  \fig{regions} (a) describes the phase space for the naive collinear contribution. If there were no jet algorithm, $p^\perp_g$ would be integrated all the way to infinity in the full range $0 \le p^-_g \le Q$. For the SW algorithm, the region defined by $p^-_g \le 2 \beta \, Q$ and  $2| p^\perp_g | / p^-_g \ge   \delta$ corresponds to a gluon being emitted outside the cone, and the triangular region at the bottom of the figure gives the cone contribution. The corresponding constraint function $F_{\mbox{\scriptsize  SW}} $ is given by
\begin{align}
\label  {eq:SWF}
        F_  { \mbox{\scriptsize  SW}} &=  \theta \left( \tan  \delta-  \frac{| p^\perp_g |}{|p^3_g|} \right) \theta \left( \tan \delta - \frac{| p^\perp_q |}{|p^3_q|} \right) 
                +  \theta \left( \frac{| p^\perp_g |}{|p^3_g|} - \tan  \delta \right)\theta \left( \beta Q - p^0_g \right) \notag \\
        &= 
                 \theta \left( y- \frac{ 4 \s }{4 \s + Q^2 \delta^2} \right) \theta \left(  \frac{ Q^2 \delta^2 }{4 \s + Q^2 \delta^2}-y \right)  
                 +  \theta \left( \frac{ 4 \s }{4 \s + Q^2 \delta^2} -y \right)    \theta \left(  2 \beta - y\right)
                 +  \mathcal{O}( \lambda, \delta, \beta) \nn \\ 
        &=
                        \theta \left( y- \frac{ 4 \s }{4 \s + Q^2 \delta^2} \right) \theta \left(  \frac{ Q^2 \delta^2 }{4 \s + Q^2 \delta^2}-y \right)  
                 + \theta \left(  2 \beta - y\right)  +  \mathcal{O}( \lambda, \delta, \beta),    
\end{align}
where \eq {deltaconditions} was used in getting to the second line, and the
first theta function in the second term is always satisfied at leading order in
the power counting. This is because $ 4s / (4 \s + Q^2 \delta^2) \sim \lambda^0
$ but $y \sim \lambda$ in the second term due to $\beta \sim \lambda$. We note that the two theta functions in the first term imply  $\theta( Q^2 \delta^2/4- \s)$ and thus limit the maximum allowed jet mass.

The zero-bin phase space is shown in \fig{regions} (b). Without the algorithm,
the integration region would extend to infinity for both $p^-_g$ and
$p^\perp_g$.  As explained in Appendix \ref{sec:zbforalg}, the zero-bin
  scaling only affects the jet algorithm through the conditions in
  \eq{ZBdeltaconditions}. This results in a zero-bin constraint function 
\begin{align}
\label  {eq:SWF0}
        F_  { \mbox{\scriptsize  0,SW}} &=  \theta \left( \tan  \delta-  \frac{| p^\perp_g |}{|p^3_g|} \right) 
                +  \theta \left( \frac{| p^\perp_g |}{|p^3_g|} - \tan  \delta \right)\theta \left( \beta Q - p^0_g \right) \notag \\
                &= 
                  \theta \left( y- \frac{ 4 \s }{ Q^2 \delta^2} \right)
                 +  \theta \left(  2 \beta - y\right)    +  \mathcal{O}( \lambda, \delta, \beta),
\end{align}
where we again used the fact that $ 4s / ( Q^2 \delta^2) \sim \lambda^0 $ to
eliminate the first theta function in the second term. (This result for the constraint function in the zero-bin region agrees with
Ref.~\cite{Ellis:2010rw}.) We note that the conditions from
\eq{ZBdeltaconditions} eliminate the second theta function in the first term.
This theta function constrained the quark to be inside the cone. Its absence
can be understood physically by remembering that in the gluon zero-bin, the
quark carries all of the label momentum and is automatically inside the cone.
In this case, there is no upper limit on the jet mass $s$.

\section{Results}
\label{sec:calculations}

\subsection{Algorithm-dependent Contribution}

We are now ready to apply our general result for the case of the SW algorithm. For both  $ y $ and  $ \s / \mu^2 $, we have used the following distribution identity
\begin{align}
        \frac{\theta(y)}{y^  { 1+ \epsilon}}
        =
                - \frac{\delta(y)}{ \epsilon}+ \left[ \frac{\theta(y)}{y}\right]_+ - \epsilon  \left[ \frac{\theta(y) \ln (y) }{ y} \right]_+ +  \mathcal{O}(  \epsilon^2),
\end{align}
where $[ \theta(y) \ln^n(y)/y]_+ $ denote plus functions, which we define such that they give zero when integrated from 0 to 1. The different possible definitions and their relationship with one another are discussed for example in Appendix B of Ref.~\cite{Ligeti:2008ac}. 

Combining \eqs  {jetfunccoll}{SWF} we find for the naive collinear contribution
\begin{align}
\label{eq:jetfuncSWcoll}
        \Delta \tilde{J}_{n}^  {\rm \, SW} (\s)= & \frac{ \alpha _s C_F }{4 \pi}  A( \epsilon) \theta \left( \delta^2  - \frac{ 4\s}{ Q^2}  \right)
                \Bigg\{ \delta(\s) \left[ \frac{2}{ \epsilon^2} 
                -\frac{4}{ \epsilon}   \left\{  \ln  2 \beta + \frac{1}{2} \ln \left( \frac{ Q^2 \delta^2}{ 4 \mu^2} \right) \right\}
                + 2  \ln^2  2 \beta - \ln^2 \left(  \frac{ Q^2 \delta^2}{ 4 \mu^2} \right) \right]
                \notag \\
                &+ \frac{4}{ \mu^2} \left[ \frac{ \mu^2 \theta(\s)}{\s}\right]_+ 
                \bigg\{
                \ln  2 \beta 
                +   \ln \left( \frac{ Q^2 \delta^2}{ 4 \mu^2} \right) 
                + \frac{ 6\s}{4\s + Q^2 \delta^2}
                \bigg\}
                -  \frac{4}{ \mu^2} \bigg[  \frac{ \mu^2 \theta(\s) \ln \left( \s/ \mu^2 \right)}{\s}\bigg]_+   \Bigg\}\notag \\
                &+\frac{ \alpha _s C_F }{4 \pi}  \theta \left( \frac{ 4\s}{ Q^2} -\delta^2  \right) 
                 \frac{4}{ \mu^2} \left[ \frac{ \mu^2 \theta(\s)}{\s}\right]_+
                 \left\{
                 \frac{3}{4}            +\ln  2 \beta 
                 \right\}.       
\end{align}
\eqs  {jetfuncZB}{SWF0} give the corresponding zero-bin contribution 
\begin{align}
\label{eq:jetfuncSWzb}
        \Delta J_{n0}^  {\rm \, SW}  (\s)= & \frac{ \alpha _s C_F }{4 \pi}  A( \epsilon) \theta \left( \delta^2  - \frac{ 4\s}{ Q^2}  \right)
                \Bigg\{ \delta(\s) \left[ \frac{2}{ \epsilon^2} 
                -\frac{4}{ \epsilon}   \left\{  \ln  2 \beta + \frac{1}{2} \ln \left( \frac{ Q^2 \delta^2}{ 4 \mu^2} \right) \right\}
                + 2  \ln^2  2 \beta - \ln^2 \left(  \frac{ Q^2 \delta^2}{ 4 \mu^2} \right) \right]
                \notag \\
                &+ \frac{4}{ \mu^2} \left[ \frac{ \mu^2 \theta(\s)}{\s}\right]_+ 
                \bigg\lbrace 
                		\ln  2 \beta 
                		+   \ln \left( \frac{ Q^2 \delta^2}{ 4 \mu^2} \right) 
                \bigg\rbrace 
                -  \frac{4}{ \mu^2} \bigg[
                	 \frac{ \mu^2 \theta(\s) \ln \left( \s/ \mu^2 \right)}{\s}\bigg]_+   
	 \Bigg\}
                \notag \\
                &+\frac{ \alpha _s C_F }{4 \pi}  \theta \left( \frac{ 4\s}{ Q^2} -\delta^2  \right) 
                \left\{
                 \frac{4}{ \mu^2} \left[ \frac{ \mu^2 \theta(\s)}{\s}\right]_+
                \bigg\lbrace 
                \ln  2 \beta 
                +   \ln \left( \frac{ Q^2 \delta^2}{ 4 \mu^2} \right) 
                \bigg\rbrace
                  - \frac{4}{ \mu^2} \bigg[ \frac{ \mu^2 \theta(\s) \ln \left( \s/ \mu^2 \right)}{\s}\bigg]_+
                  \right\}.              
\end{align}
The total algorithm-dependent part  \( \Delta J_{n} \) is given by the difference of \eqs{jetfuncSWcoll}{jetfuncSWzb} and reads
\begin{align}
\label{eq:jetfuncSWfull}
        \Delta J_{n}^  {\rm \, SW}  (\s)= & \frac{ \alpha _s C_F }{4 \pi}\theta \left( \delta^2 - \frac{ 4\s}{ Q^2}  \right)
                \frac{24}{ 4\s + Q^2  \delta^2}
                +\frac{ \alpha _s C_F }{4 \pi}  \theta \left( \frac{ 4\s}{ Q^2} -\delta^2  \right) 
                 \left\{
                \frac{3}{ \s} 
                  +   \frac{4}{ \s} \ln \left( \frac{ 4 \s}{ Q^2 \delta^2} \right) 
                  \right\}.      
\end{align}
Using the definition of the plus function \cite{Ligeti:2008ac}, it can be seen that when $[ \theta(y) \ln^n(y)/y]_+ $ is integrated against a function $f(y)$ such that $f(0)=0$, we can ignore the plus function prescription because it only makes a difference at $y=0$. Hence, we do not need to use plus functions in \eq{jetfuncSWfull}. We note that the outside-cone region for the gluon cancels between the collinear and the  zero-bin contributions and hence there is no \(\beta\)-dependence in the final result. This cancellation has been discussed in the literature \cite{Cheung:2009sg} and has to take place since the purpose of the zero-bin subtraction is to remove any soft contribution from the collinear diagrams and the gluon must be soft to be outside the cone. We have also grouped the terms according to whether  $ \s $ is limited to be less or greater than  $ (Q \delta /2)^2 $.

\subsection{Full Sterman-Weinberg Jet Function And Anomalous Dimension}
\begin{figure} 
\begin{center}
\includegraphics[width=3in]{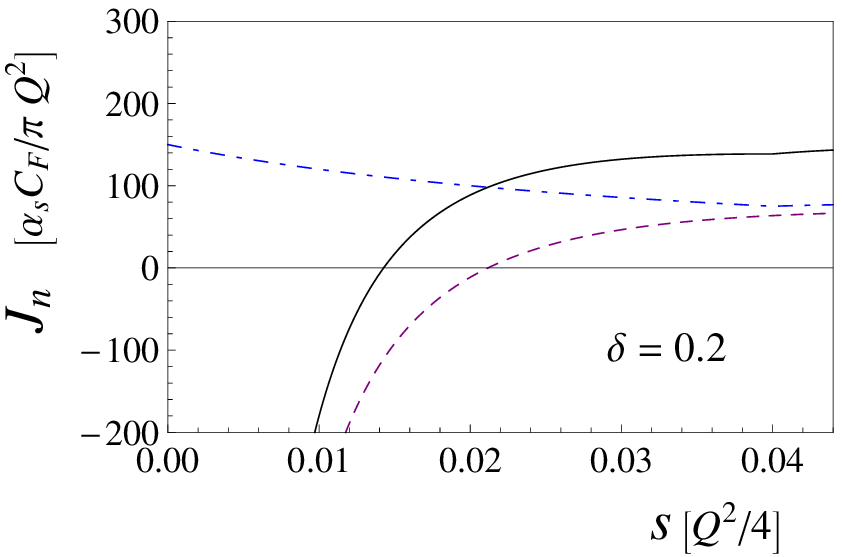} \hspace{3em}
\includegraphics[width=3in]{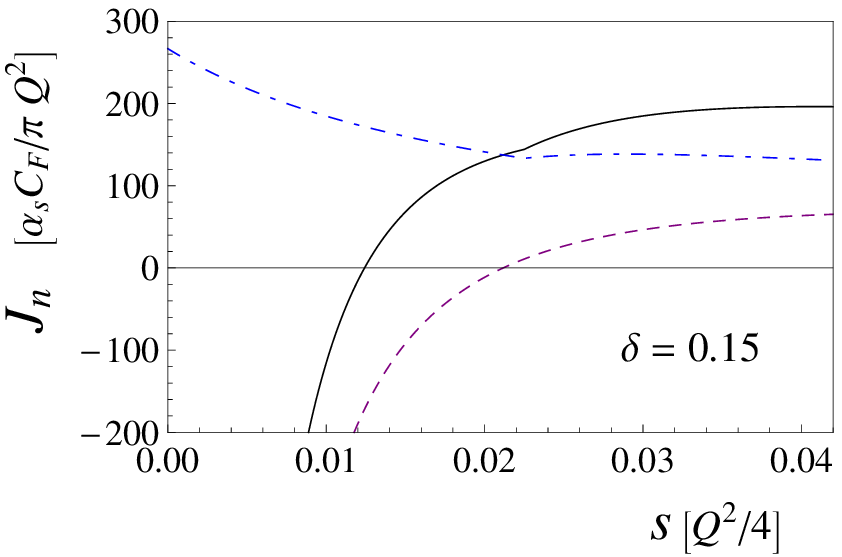}
\caption{The renormalized jet function in units of $\alpha_s(\mu) C_F/\pi Q^2$, where we choose $\mu = 0.1 \times Q/2$. The SW (solid black line) and the inclusive (dashed purple line) jet functions are plotted as well as the difference of the two (dash-dotted blue line) for $ \delta = 0.2 $ (left) and $ \delta = 0.15 $ (right). 
}
\label{fig:JPlot}
\end{center}
\end{figure}

We will now give an explicit expression for the full SW jet function. To do so, we need the inclusive jet function $J^{\rm (inc)}_{n}$, which can be found in the literature \cite{Bauer:2003pi, Becher:2006qw}. Using the notation of Ref.~\cite{Fleming:2007xt}, it can be written as
\begin{align}
\label{eq:jetfuncInclfull}
         J^{\rm (inc)}_{n} (\s,\mu)= \delta(\s)  + \frac{ \alpha _s C_F }{4 \pi}
                \Bigg\{
                 \delta(\s) \left( \frac{4}{ \epsilon^2} + \frac{3}{ \epsilon}+7- \pi^2 \right)- \frac{4}{ \mu^2} \left[ \frac{ \mu^2 \theta(\s)}{\s}\right]_+
                 \left(
                 \frac{1}{ \epsilon}+\frac{3}{4}  \right)
                  + \frac{4}{ \mu^2} \bigg[
                  	\frac{ \mu^2 \theta(\s) \ln \left( \s/ \mu^2 \right)}{\s}\bigg]_+
                  \Bigg\}.      
\end{align}
The SW jet function is given by the sum of \eqs{jetfuncSWfull}{jetfuncInclfull}. Since the jet algorithm does not modify the $1/ \epsilon$-poles, the anomalous dimension is
\begin{align} \label{eq:gammaJn}
  \gamma^  {\rm \, SW}_{J_n}(s, \mu) & = -  \frac{\alpha_s C_F}{4\pi}\bigg\{
   \frac{8}{ \mu^2} \bigg[ 
   \frac{ \mu^2\theta(\s )}{\s} \bigg]_+
   - 6\,  \delta(\s)
   \bigg\} .
   \end{align}
The renormalized SW jet function in $ \overline  { \text{MS}} $-scheme is given by
\begin{align} \label{eq:renJn} 
J^  {\rm SW}_{n,\rm{ren}}(\s, \mu)  &= \delta(s)  + \frac{\alpha _s C_F}{4\pi} \Bigg\{ \delta(s)
        \left(
         7 - \pi^2 \right)
        -  \frac{3}{\mu ^2}\bigg[ \frac{\mu ^2 \theta(s)}{s}\bigg ]_+ %
        + \frac{4}{\mu ^2} \bigg[ \frac{\mu ^2 \theta(s) \ln(s/\mu ^2)}{s}\bigg]_+ \nn \\
        &+\theta \left( \delta^2 - \frac{ 4\s}{ Q^2}  \right)
                \frac{24}{ 4\s + Q^2  \delta^2}
                +\theta \left( \frac{ 4\s}{ Q^2} -\delta^2  \right) 
                 \left[
                \frac{3}{ \s} 
                  +   \frac{4}{ \s} \ln \left( \frac{ 4 \s}{ Q^2 \delta^2} \right) \right]
 \Bigg\}
\end{align}
The most important result of this work is the
renormalized SW jet function in \eq{renJn}. It is
plotted in \fig{JPlot} for $\delta=0.2$ and $\delta=0.15$ while keeping $ s \ne
0 $. We have chosen the jet scale to be $ \mu = 0.1 \times(Q/2)$. The SW and the
inclusive jet functions are denoted by the solid black line and the dashed
purple line, respectively. The jet function is continuous, but the derivative of the algorithm dependent contribution given in \eq{jetfuncSWfull} changes sign at  $\s = (Q \delta /2)^2 $,  where one theta function turns off and the other turns on. We note that the algorithm contribution denoted by the dash-dotted blue
line is always positive. 

Comparing the algorithm contributions in \fig{JPlot} for the two values of the cone angle, one can see a change in shape. For $\delta = 0.15$, small values of $s$ --- which correspond to narrow jets --- contribute more than large values of $s$. For $\delta = 0.2$, the algorithm contribution is flatter, signifying a more equal contribution from jets of different size. This makes physical sense since a broader cone allows contributions from wider jets. Unlike the shape, the overall normalization of the algorithm contribution does not follow from physical intuition --- we might have naively expected that increasing the cone angle would give a larger value for the magnitude of the jet function. However, care must be taken in assigning physical meaning to the normalization of a jet function computed in an unphysical subtraction scheme ($\overline {\rm MS}$), which is also illustrated by the fact that the inclusive jet-function has a zero at finite $s$. Different renormalization schemes include different constant pieces in the jet function and make its magnitude scheme dependent. Additionally, it should be remembered that a jet algorithm also affects the soft function, which must be combined with jet functions to see the full $\delta$ dependence of the cross section.
\begin{figure} 
\begin{center}
\includegraphics[width=3in]{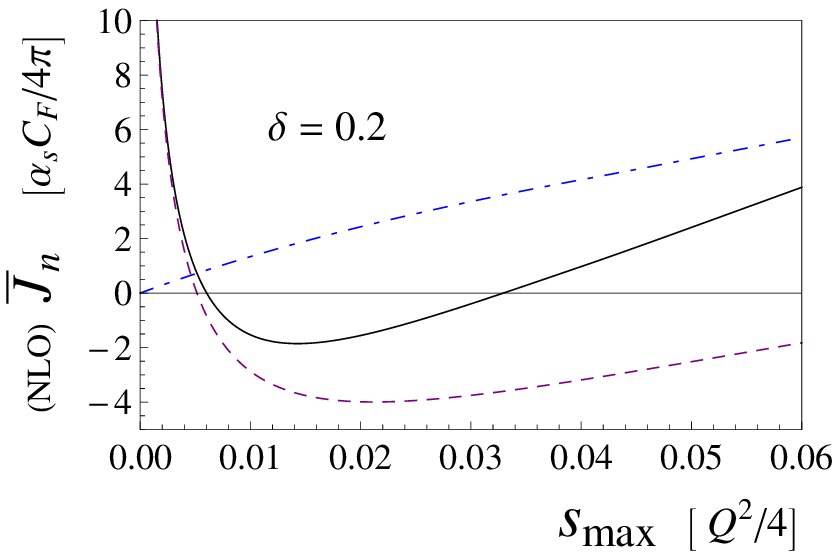} \hspace{3em}
\includegraphics[width=3in]{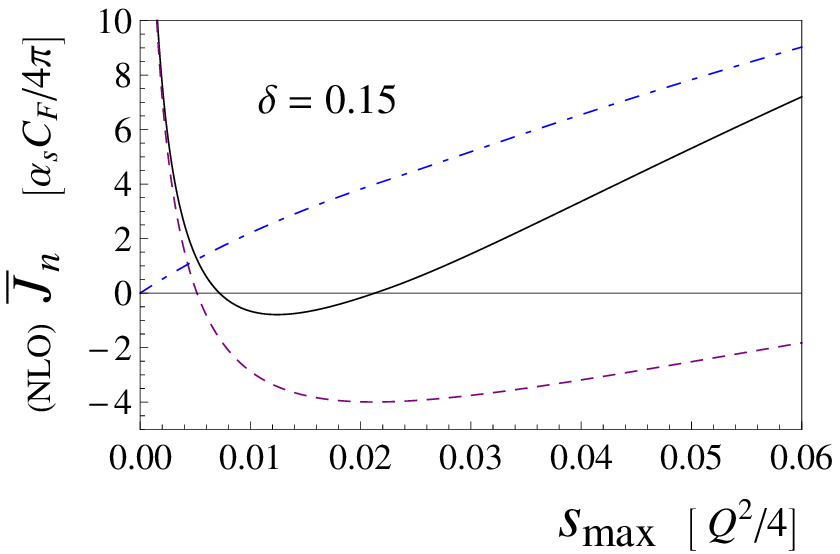} \newline \newline \newline
\includegraphics[width=3in]{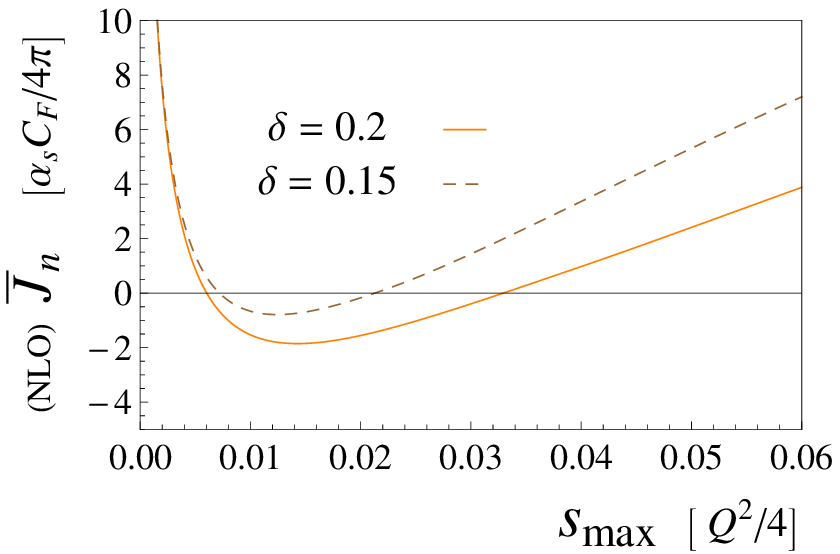} 
\caption{The renormalized NLO contribution to the integrated jet function is plotted in units of $\alpha_s(\mu) C_F/4\pi$, where we choose $\mu = 0.1 \times Q/2$. On the first row we plot the SW (solid black line) and the inclusive (dashed purple line) integrated jet functions as well as the difference of the two (dash-dotted blue line) for $ \delta = 0.2 $ (left) and $ \delta = 0.15 $ (right). On the second row we compare the integrated SW jet functions for the two values of $\delta$. }
\label{fig:IntJPlot}
\end{center}
\end{figure}

It is also interesting to integrate \eq{renJn} over $\s$ up to $\s_{\rm max}$. We call this the integrated jet function $\bar{J}^{\rm \, SW(\s_{\rm max})}_{n,\rm{ren}}(\mu)$ and it is given by
\begin{align} \label{eq:IntjetfuncSW}
&\bar{J}^{\rm \, SW (\s_{\rm max})}_{n,\rm{ren}}(\mu)\equiv \int^{\s_{\rm max}}_0 \df s \, J^{\rm \, SW}_{n,\rm{ren}}(\s,\mu)\notag \\
&=1+ \theta \left( \delta^2  - \frac{ 4 \s_{\rm max}}{ Q^2}  \right)
        \frac{ \alpha _s C_F }{4 \pi}
                \Bigg\{
                  7- \pi^2 
                  -  3 \ln \left( \frac{  \s_{\rm max}}{\mu^2}\right)           
                  +2 \ln^2 \left( \frac{  \s_{\rm max}}{\mu^2}\right) + 6 \ln \left( \frac{  4 \s_{\rm max}+Q^2 \delta^2}{Q^2 \delta^2}\right) 
                  \Bigg\}
                   \notag \\
                   &\phantom{=}
                   +
                   \theta \left( \frac{ 4 \s_{\rm max}}{ Q^2}- \delta^2  \right)
                \frac{ \alpha _s C_F }{4 \pi}
                \Bigg\{
                  7- \pi^2 
                  -  3 \ln  \left( \frac{ Q^2 \delta^2 }{4 \mu^2}\right)        
                  +2 \ln^2  \left( \frac{ Q^2 \delta^2 }{4 \mu^2}\right)  + 6 \ln 2 
                  + 4 \ln \left( \frac{  \s_{\rm max}}{\mu^2}\right) \ln \left( \frac{ 4 \s_{\rm max} }{ Q^2 \delta^2}\right)
                  \Bigg\}.
\end{align}
The renormalized NLO contribution to the integrated jet function is plotted in \fig{IntJPlot} as a function of $\s_{\rm max}$ and we have again chosen the jet scale to be $ \mu = 0.1 \times(Q/2)$. On the first row of \fig{IntJPlot}, we plot the jet function for the SW case and
the inclusive case, as well as the difference of the two, for $ \delta=0.2$ and
$\delta = 0.15 $. Since the difference between the SW and the inclusive jet
functions was always positive in \fig{JPlot}, we can see that the difference
between the integrated SW and inclusive jet functions in \fig{IntJPlot}
increases monotonically as a function of $\s_{\rm max}$. On the second row of
\fig{IntJPlot}, the integrated SW jet functions for the two different values of
$\delta$ are compared.

\section{Conclusions}
\label{sec:conclusions}

Using a generic jet algorithm, we have derived an expression for the jet
function \( J^F_{n}(\s,\mu) \) as a function of the jet invariant mass $ \s$,
valid up to \( \mathcal{O}(\alpha_s) \) and to leading order in the power
counting. Expressions for the naive collinear and the zero-bin contributions
have been given. We have demonstrated the general result by calculating the jet
function together with its anomalous dimension for the Sterman-Weinberg
algorithm, which is parameterized by the cone half-angle $\delta$ and the energy
fraction of the soft radiation $\beta$. The anomalous dimension of the jet function is not modified by the jet
algorithm. It was shown that all \(\beta\) dependence is canceled because
the gluon outside the cone must be soft and hence its contribution is removed by
the nontrivial zero-bin subtraction.  Our result for the anomalous dimension 
agrees with that reported in Ref.~\cite{Ellis:2009wj}. We have also calculated
the finite part of the jet function.

The renormalized jet function has been plotted as a function of the jet
invariant mass to illustrate the difference between the SW and the inclusive
case.  We have also defined the integrated SW jet function \( \bar{J}^{\rm \, SW
  (\s_{\rm max})}_{n}(\mu) \) by integrating \( J^ {\rm SW}_{n}(\s, \mu) \) up
to $\s_{\rm max}$.  We have shown how \( \bar{J}^{\rm \, SW}_{n}(\mu) \) changes
as a function of $\delta$ and how it differs from the inclusive case.

After the soft function for the SW algorithm has been calculated, our result for the jet function can be used to derive a jet algorithm dependent factorized cross section for $ e^+e^- $ into dijets  as a function of the jet invariant mass  $ \s $.
\newline \newline
\emph{Note added}: Right after this paper first appeared, Ref.~\cite{Ellis:2010rw} by Ellis et al.~came out with a calculation of the full jet
function. They pointed out that the treatment of the zero-bin was different in our two papers. In Appendix \ref{sec:zbforalg}, we
derive a systematic method to apply zero-bin scaling to phase space constraints. With this method we find the same zero-bin subtractions as they do. After a private communication to sort out a typographical error in the theta functions in their original manuscript, the results for the full jet function agree between the two papers.

\begin{acknowledgments}
  I am very grateful to Iain Stewart for his guidance during this project and
  for feedback on the manuscript. I would like to thank Frank Tackmann and
  Carola Berger for their help with the calculations and their comments on the
  manuscript, and Claudio Marcantonini and Wouter Waalewijn for useful discussions.
  I also wish to thank the authors of Ref.~\cite{Ellis:2010rw} for their comments on
  zero-bin subtractions. This work was supported in part by the Office of
  Nuclear Physics of the U.S. Department of Energy under the Contract
  DE-FG02-94ER40818.
\end{acknowledgments}

\appendix
\section{Zero-bin with a Jet Algorithm}
\label{sec:zbforalg}

The purpose of the zero-bin subtraction is to remove double counting between the
jet and soft functions. As explained in Ref.~\cite{Manohar:2006nz}, there is some
freedom in how to define the subtraction. This can be compared to freedom in choosing from different renormalization schemes, all of which remove the UV divergences but can differ by finite constants. Similarly, all zero-bin subtraction schemes must remove the IR divergences in a universal manner but may include different constants in the result.  The authors of Ref.~\cite{Manohar:2006nz} advocate for a minimal approach analogous to minimal subtraction for renormalization. We show how their approach can be extended in a consistent way to apply also in the presence of phase space restrictions, such as jet algorithms. We expect there to be other consistent zero-bin subtraction schemes but we leave their exploration to future work. 

When dealing with phase space, it is convenient to think about the
zero-bin scaling in terms of final state momenta instead of loop momenta. The
two approaches are equivalent but the former is conceptually simpler to apply to
jet algorithms. To begin with, we write down the Feynman diagrams for the desired process where
an initial parton goes into a final state of several particles.  The momenta of
all the external particles and all the internal propagators are considered
independent and momentum conservation at the vertices is implemented by explicit
delta functions and integrals over the internal momenta. At any order in
$\alpha_s$, the zero-bin contributions for a Feynman diagram are found by taking
one or more of the collinear final state momenta $ p_i $ to scale as $p_i \sim Q
\lambda^2 $. This only affects expressions where two different momenta are added
or subtracted. In the approach advocated here such comparisons only take place at the vertices. 
Performing the integrals over the
internal momenta conveys the information about the zero-bin scaling contained in
the vertices into the rest of the diagram. The same approach can also be
taken when thinking about the zero-bin scaling in loop diagrams that give the
cross section via the optical theorem. As long as the momentum for every
propagator is considered independent, it is possible to apply the zero-bin
scaling to a single propagator and let the momentum-conserving delta functions
convey the information about the scaling to the rest of the propagators. Hence,
thinking about the zero-bin this way unifies the treatment of phase space
integrals and loop integrals.
\begin{figure}
       \begin{center}
       \includegraphics[width=2.5in]{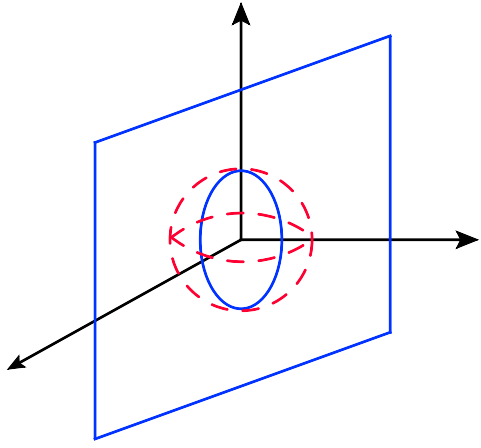} 
       \caption{Schematic representation of the phase space. The blue parallelogram represents the hypersurface defined by phase space constraints, the red dashed sphere represents the zero-bin region, and the blue oval shows the intersection of the hypersurface with the zero-bin region.}
       \label{fig:phasespace}
       \end{center}
\end{figure} 

In order to discuss the influence of phase space restrictions, let us consider a
cross section that is fully differential in $N$ variables, i.e.~no integrals
over physical degrees of freedom have been performed. In this case, no zero-bin
subtraction is needed since we have full control over the momenta of all
particles; we can tell which region of phase space they are in and hence
whether they are soft or collinear.  Next, let us perform some integrals that
can cause the integrand to be evaluated in the zero-bin region while keeping the
cross section differential in $M$ variables $b_k$. (For example, think of $b_k$
as the invariant masses of the jets.)  In this case, the integration region
becomes a hypersurface in phase space, represented schematically in
\fig{phasespace} by the blue parallelogram. The red dashed sphere represents the
zero-bin region and the blue oval is the intersection of the zero-bin region
with the hypersurface on which we have restricted the final state particles. The
region inside the blue oval has to be removed by the zero-bin subtraction but
there is no need to perform a subtraction outside the hypersurface. This is
another way of saying that the zero-bin scaling does not act directly on the
phase space constraints that specify the values for the $b_k$. However, the
constraints can be modified indirectly by the momentum-conserving delta
functions.

To state the argument more mathematically, consider two ways to perform the phase space integrals over the hypersurface. Either we integrate over all $N$ degrees of freedom and use delta functions to enforce the $M$ constraints or alternatively, we use the constraints to find a set of $N-M$ independent coordinates  $ q_i $ and integrate over them. The two approaches can be written as
\begin{equation} \label{eq:phasespace}
\int \prod_{i=1}^N \df p_i \prod_{k=1}^M \delta (b_k - \hat{b}_k) = \int \prod_{i=1}^{N-M} \df q_i .
\end{equation}
If we use the right hand side of \eq  {phasespace} to evaluate the cross section, we can see that the zero-bin scaling will be applied to the integrand but will not affect the phase space constraints  $ \delta (b_k - \hat{b}_k) $ directly. Similarly, we can apply the phase space constraints required by a jet algorithm so that the integration is performed over a region of phase space. Again, the zero-bin subtraction has to be performed on phase space region specified by the algorithm and the algorithm constraints are only modified by the effect of the zero-bin scaling on the momentum-conserving delta functions.

To give a concrete example, we look at the zero-bin subtraction for the diagrams in \fig{feyndiag}. The correspondence between the original
momentum-conserving delta functions and the associated zero-bin scaled versions is
\begin{align}
\label{eq:ZBscaling}
        \delta \left( Q-p^-_q -p^-_g \right) &\rightarrow \delta \left( Q-p^-_q \right) \nn \\
        \delta^  {d-2} \left(p^\perp_q -p^\perp_g \right) &\rightarrow \delta^  {d-2} \left(p^\perp_q \right) \nn \\
        \delta \left( \s/ Q-p^+_q -p^+_g \right) &\rightarrow \delta \left( \s/ Q-p^+_q -p^+_g \right).
\end{align}
Applying these delta functions to the on-shell condition for the quark gives
\begin{align}
\label{eq:ZBonshell}
        \delta (p^-_q p^+_q + \left( p^\perp_q  \right)^2) &\rightarrow \delta (Q p^+_q ).
\end{align}
Combining \eqs {ZBscaling}{ZBonshell} with the gluon on-shell condition leads to
\eq{ZBdeltaconditions}, which together with \eq{SWF} gives \eq{SWF0}. 

It should be cautioned that performing the zero-bin subtraction as explained
above does not give the same result as  applying the scaling $ p_g \sim Q \lambda^2 $ directly to
the jet algorithm constraints.  In the latter approach, the gluon angle with respect to the jet axis would scale as \( \lambda^0 \) and the theta function would never be satisfied according to power counting since \(\delta \sim \lambda\).

In Ref.~\cite{Ellis:2010rw}, Ellis et al.~point out that the difference between the results in our two papers could come
from a different treatment of the zero-bin and from the different power counting
that we use for $\delta$.  Our calculation corresponds to their measured jet
function with angularity $\tau_0 = \s/Q^2$ and cone half-angle $R = \delta$.
They use power counting $\tau_0/R \sim \lambda^2$ whereas in the present work we
take $\tau_0 \sim R \sim \lambda^2$. The power counting of momenta is the same
in both works. We carried out the investigations reported in this appendix in
order to clarify these differences. Following the approach advocated above, we
derive the same zero-bin subtractions as Ref.~\cite{Ellis:2010rw}, indicating
that both calculations use minimal zero-bin subtractions rather than some other
scheme. 

On the question of power counting of  $ \delta $, if we write $\delta \approx 2 \tan
\delta/2$, it can be seen that at leading order in power counting of momenta our
jet algorithm constraints in \eqs{SWF}{SWF0} are equivalent to their Eqs.~(5.1)
and (5.3).  However, the terms appearing in their
theta functions in Eq.~(5.1) are of different order in \(\lambda\), whereas our
theta functions constraining the naive contribution in \eq{SWF} are homogeneous in power counting. For the phase space constraints in the zero-bin region, the scaling will not be homogeneous because, as argued above, we do not need to apply the zero-bin scaling to the constraints. Because Ellis et al.~do not expand in \(\lambda\) in the theta functions for the naive contribution, our algorithm constraints end
up being identical to theirs and our \eqs{jetfunccoll}{jetfuncZB} are identical to their
Eqs.~(5.7) and (5.8) with a cone algorithm. 

The remaining difference between the two papers in the finite part of the jet function is due to a typographical error in Ref.~\cite{Ellis:2010rw} where some of the theta functions have been misplaced.
\footnote  {In a private communication, the authors of Ref.~\cite{Ellis:2010rw} confirmed the presence of a typographical error in the first version of their paper. }
With this fix in the theta functions, our two papers give the same result for the jet function.


\bibliographystyle{physrev4}
\bibliography{/Users/teppo/Documents/Bibliographies/MainBibliography}

\end{document}